\documentclass[12pt,letter]{iopart}
\usepackage{graphicx}
\usepackage{setspace}
\usepackage{texdraw}
\usepackage{amsfonts}

\newcommand\T{\rule{0pt}{7ex}}
\newcommand\B{\rule[-5.5ex]{0pt}{0pt}}
\newcommand\TBT{\rule{0pt}{6ex}}
\newcommand\BTB{\rule[-4.5ex]{0pt}{0pt}}
\newcommand\TT{\rule{0pt}{4.5ex}}
\newcommand\BB{\rule[-3ex]{0pt}{0pt}}

\begin{document}
\title{Ellipses of Constant Entropy in the $XY$ Spin Chain}

\author{F. Franchini $\star$,  \ A. R. Its\dag,  \ B.-Q. Jin\S   \ and V. E. Korepin \maltese }

\address{ $\star$ Department of Physics and Astronomy, State University of New York at \\
Stony Brook, Stony Brook, NY 11794, USA}
\address{ \dag\ Department of Mathematical Sciences, Indiana
University-Purdue University Indianapolis, Indianapolis, IN
46202-3216, USA}
\address{\S\ College of Physics and Electronic Information, Wenzhou
University, Wenzhou, Zhejiang, P.R. China }
\address{$\maltese$ \ C.N.\ Yang Institute for Theoretical Physics, State
University of New York at Stony Brook, Stony Brook, NY 11794-3840,
USA}

\ead{fabio.franchini@stonybrook.edu,itsa@math.iupui.edu,jinbq@wzu.edu.cn,
korepin@insti.physics.sunysb.edu}


\begin{abstract}
Entanglement in the ground state of the  $XY$ model on the
infinite chain can be measured by the von Neumann entropy of a
block of neighboring spins. We study a double scaling limit: the
size of the block is much larger than $1$ but much smaller than
the length of the whole chain. The entropy of the block has an
asymptotic limit. We study this limiting entropy as a function of
the anisotropy and of the magnetic field. We identify its minima
at product states and its divergencies at the quantum phase
transitions. We find that the curves of constant entropy are
ellipses and hyperbolas and that they all meet at one  point ({\it
essential critical point}). Depending on the approach to the
essential critical point the entropy can take any value between
$0$ and $\infty$. In the vicinity of this point small changes in
the parameters cause large change of the entropy.
\end{abstract}

\maketitle

\section{Introduction}

Entanglement is a primary resource for quantum information
processing \cite{BD,L, ben,pop}. It shows how much quantum effects
we can use to control one system by another. Stable and large
scale entanglement is necessary for scalability of quantum
computation \cite{rasetti,GRAC}. The entropy of a subsystem as a
measure of entanglement was discovered in \cite{ben}. Essential
progress has been achieved in the understanding of entanglement in
various quantum systems
\cite{rasetti,eisert,eisert1,eisert2,eisert4,eisert5,plenio,brandt,fazio,amico,jin,LRV,K,julien,
cardy,salerno,ABV,VMC,LO,PP,fan,eisert3,briegel,bruno}.

The importance of the $XY$ model for quantum information  was
emphasized in \cite{fazio,vidal,keat,bose}\footnote{It is
interesting to note that the critical behavior of the $XY$ model
is similar to the Lipkin-Meshkov-Glick model \cite{julien}.}. In
this paper we consider the entropy of a block of $L$ neighboring
spins in the ground state of the $XY$ model [on the infinite
chain] in the limit $L\rightarrow \infty $. We use the results of
\cite{bik, pesh, big} and extend them to the whole phase diagram
of the model.\footnote{Moreover, in the Appendix we explain how
these results can be used to calculate the entanglement of the
$XY$ model in a staggered magnetic field.} The Hamiltonian of the
$XY$ model is
\begin{equation}
  {\cal H}=-\sum_{n=-\infty}^{\infty}
  (1+\gamma)\sigma^x_{n}\sigma^x_{n+1}+(1-\gamma)\sigma^y_{n}\sigma^y_{n+1} +
  h \; \sigma^z_{n}
  \label{xxh}
\end{equation}
Here $\gamma \ge 0$ is the anisotropy parameter; $\sigma^x_n$,
$\sigma^y_n$ and $\sigma^z_n$ are the Pauli matrices and $h \ge 0$
is the magnetic field. The model is clearly symmetric under the
transformation $\gamma \to - \gamma$ or $h \to -h$. In
\cite{bik,big} only the case $0 \le \gamma \le 1$ was discussed,
here we can confirm that those results can be directly extended
for $\gamma > 1$ by analytical continuation.

The $XY$ model was solved in \cite{Lieb,mccoy,mccoy2,gallavotti}.
The methods of Toeplitz determinants and integrable Fredholm
operators were used for the evaluation of correlation functions,
see \cite{mccoy2,aban,sla, dz, izer, pron}. The idea to use the
determinants for the calculation of the entropy was put forward in
\cite{jin}.

The solution of the $XY$ model looks  differently in {\bf three
cases:}

{\sc Case $2$} is defined by $h > 2$:
  This is strong magnetic field.

 {\sc Case $1$a} is defined by $h < 2$ and $\gamma > \sqrt{1-(h/2)^2}$:
 Moderate magnetic field for small anisotropy, and includes zero magnetic field for large anisotropy.

 {\sc Case $1$b} is defined by $h < 2$ and $\gamma < \sqrt{1- (h/2)^2}$:
 It describes weak magnetic field, including zero magnetic field in the small anisotropy regime.

\begin{figure}
\begin{center}
   \dimen0=\textwidth
   \advance\dimen0 by -\columnsep
   \divide\dimen0 by 2
\includegraphics[width=\columnwidth]{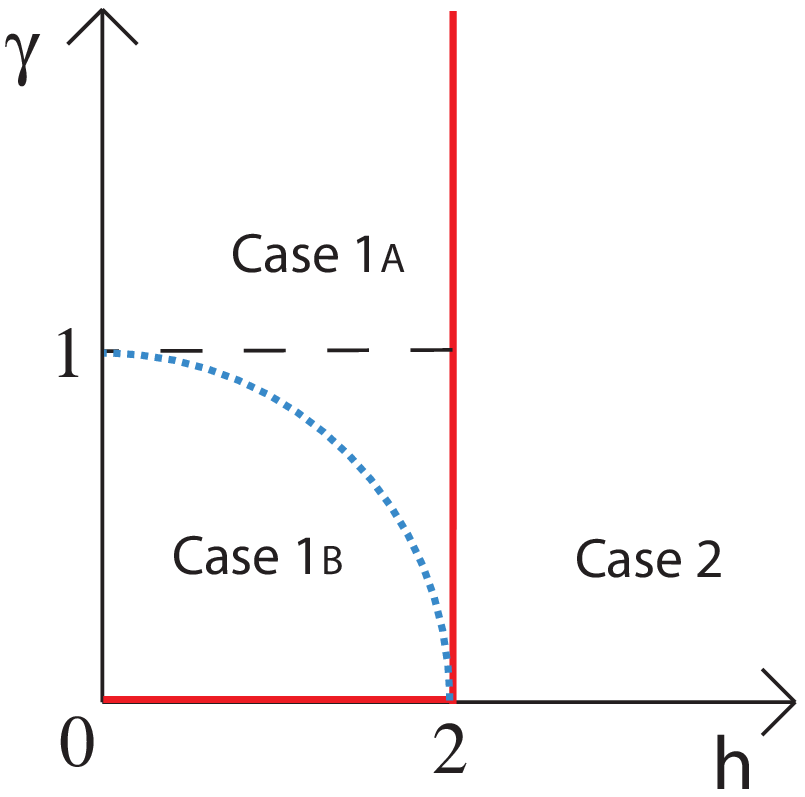}
\caption {Phase diagram of the anisotropic $XY$ model in a
constant magnetic field (only $\gamma \ge 0$ and $h \ge 0$ shown).
The three cases {\sc $2$, $1$a, $1$b}, considered in this paper,
are clearly marked. The critical phases ($\gamma = 0$, $h \le 2$
and $h = 2$) are drawn in bold lines (red, online). The boundary
between cases {\sc $1$a} and {\sc $1$b}, where the ground state is
given by two degenerate product states, is shown as a dotted line
(blue, online). The Ising case ($\gamma = 1$) is also indicated,
as a dashed line.}
   \label{phasediagram}
\end{center}
\end{figure}

At $\gamma = 0$ and for $h \le 2$ the model is known as the
isotropic $XY$ model (or XX model) and its spectrum is gapless.
The entropy for this critical phase was calculated in \cite{jin}.
The boundary between cases $2$ and {\sc $1$a} ($h = 2$) is also
critical. In the rest phase diagram, the spectrum of the $XY$
model is given by:
\begin{equation}
   \epsilon_k = \sqrt{ \left( \cos k - h/2 \right)^2 + \gamma^2 \; \sin^2 k } .
\end{equation}
We draw the phase diagram and the three cases we are considering
in Fig.~\ref{phasediagram}.

At the boundary between cases {\sc $1$a} and {\sc $1$b} ($
h=2{\sqrt{1-\gamma^2}} $) the ground state can be expressed as a
{\it product state} as it was discovered in \cite{shrock}. The
ground state is in fact {\bf doubly degenerated}:
\begin{eqnarray}
   |GS_1\rangle & = & \prod_{n\in \mbox{\rm lattice}}\left[ \cos(\theta)| \uparrow _n \rangle + \sin(\theta)| \downarrow _n \rangle \right] ,
  \nonumber   \\
  |GS_2\rangle & = &  \prod_{n\in \mbox{\rm lattice}}\left[ \cos(\theta)| \uparrow _n\rangle - \sin(\theta)| \downarrow _n \rangle \right]
  \label{deg}
\end{eqnarray}
Here  $\cos^2 (2 \theta) =(1-\gamma)/(1+\gamma)$. The role of
factorized states was emphasized in \cite{aban, ver,tog}. Let us
mention that the rest of the energy levels are separated by a gap
and correlation functions decay exponentially. The boundary
between cases {\sc $1$a} and {\sc $1$b} is not a phase transition.

\section{Block Entropy}

In general, we denote  the ground state of the model by
$|GS\rangle$. We consider the entropy of  a block  of $L$
neighboring spins: it measures the entanglement between the block
and the rest of the chain \cite{ben,vidal}. We treat the whole
ground state as a binary system $|GS\rangle = |A \& B\rangle $.
The block of $L$ neighboring spins is subsystem A and the rest of
the ground state is subsystem B. The density matrix of the ground
state is \mbox{$\rho_{AB}=|GS\rangle \langle GS|$}. The density
matrix of the block is \mbox{$\rho_A= Tr_B(\rho_{AB})$}. The
entropy $S(\rho_A)$ of the block is:
\begin{equation}
S(\rho_A)=-Tr_A(\rho_A \ln \rho_A)   \qquad \qquad \qquad  \sharp  \label{edif}
\end{equation}
{\bf Note that each of the ground states (\ref{deg}) is factorized and
has no entropy.}

To express the entropy we need the complete elliptic integral of
the first kind,
\begin{equation}
I(k) = \int_{0}^{1}\frac{dx}{\sqrt{(1-x^2)(1 - k^{2}x^{2})}}  \qquad \qquad \star
\nonumber
\end{equation}
and the modulus
\begin{equation}
   \tau_0= I(k')/I(k), \qquad \qquad k'=\sqrt{1-k^2}
\end{equation}

The magnetic field and anisotropy define the elliptic parameter
$k$:
\begin{eqnarray}
   k= \left \{ \begin {array} {ll}
   \gamma\; / \;\sqrt{(h/2)^2+\gamma^2-1}, &  \mbox{\sc Case $2$} \\ [0.3cm]
   \sqrt{(h/2)^2+\gamma^2-1}\; /\; \gamma ,  & \mbox{\sc Case $1$a} \\ [0.3cm]
   \sqrt{1 -\gamma^2 - (h/2)^2}\; / \;\sqrt{1-(h/2)^2}, & \mbox{\sc Case $1$b}
 \end{array}
  \right.
    \label{modMay}
  \end{eqnarray}
$k$ vanishes at large magnetic fields ($h \to \infty$), at
$\gamma=0$ for $h>2$ and at the boundary between case {\sc $1$a}
and {\sc $1$b} ($h = 2 \sqrt{1 - \gamma^2}$). In all these regions
of the phase diagram, the ground state of the system is given by
product states (a ferromagnetic state in the first two cases, the
doubly degenerate state (\ref{deg}) for the latter). At the phase
transitions ($h=2$ and $\gamma=0$, $h<2$) the elliptic parameter
$k=1$.

\begin{figure}
\begin{center}
\includegraphics[width=\columnwidth]{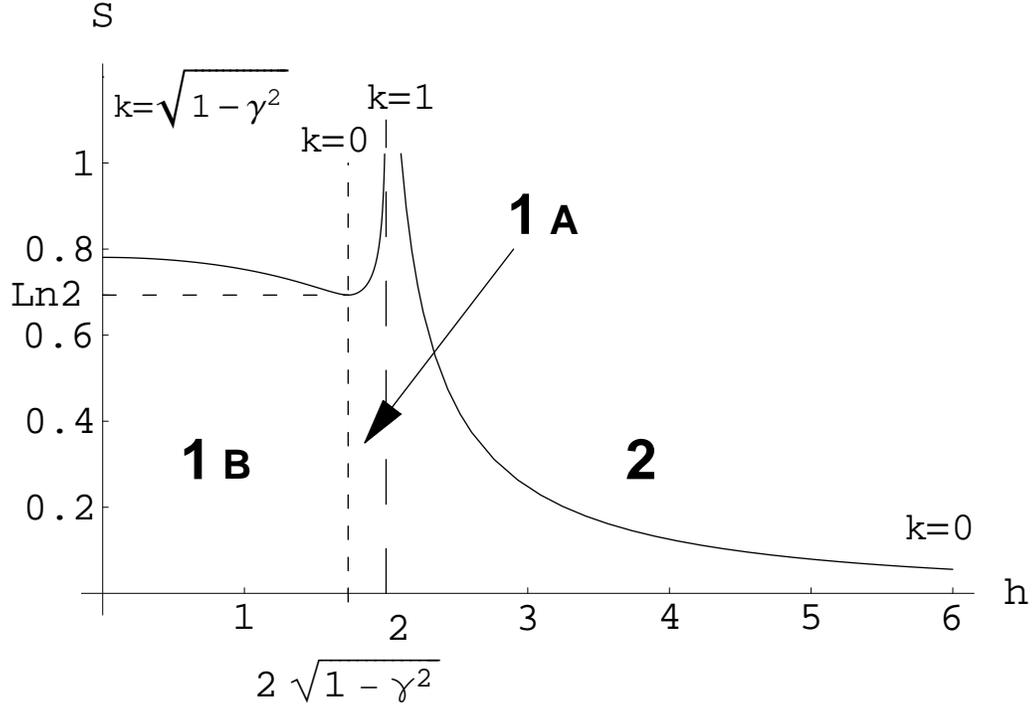}
\caption {The limiting entropy as a function of the magnetic field
at constant anisotropy $\gamma = 1/2$. The entropy has a local
minimum $S=\ln2$ at $h=2\sqrt{1-\gamma^2} $ and the absolute
minimum for $h \to \infty$ where it vanishes. $S$ is singular at
the phase transition $h=2$ where it diverges to $+ \infty$. The
three cases are marked.}
   \label{entropyplot}
\end{center}
\end{figure}

In the paper \cite{bik} we used determinant representation for the
evaluation of the entropy. The zeros of the determinant form an
infinite sequence of  numbers:
\begin{equation}
\label{zerosMay}
\lambda_{m} =
  \tanh \left(m + \frac{1-\sigma}{2}\right)\pi \tau_{0}
\end{equation}
here $\sigma = 1$ in Case 1 and $\sigma = 0$ in Case 2.
Note
 $0 < \lambda_{m} < 1$ and  $ \lambda_{m} \to 1$ as  $ m \to \infty$

These zeros allowed us to represent the entropy as a convergent
series in \cite{bik}:
\begin{equation}
\label{3333May}
 \heartsuit  \quad \qquad  S(\rho_A) =
  \sum_{m=-\infty}^{\infty}
  (1+\lambda_{m})\ln \frac{2}{1+\lambda_{m}}  \qquad  \spadesuit
\end{equation}
I. Peshel  also obtained the series (\ref{3333May}) in cases of
non-zero magnetic field, see  \cite{pesh}.\footnote{In comparing
with the results of \cite{pesh}, the reader should keep in mind
that Peshel calculated the entropy per boundary, therefore his
results differ by a factor of 2 compared to the ones in this
paper.} He summed it up into:
\begin{eqnarray}
   {\hskip -1cm } S(\rho_A) =  \frac {1} {6} \left [\;\ln{ \frac {4} {k \; k'}} + (k^2-k'^2)
   \frac {2 I(k) I(k')} {\pi} \right ],
   & \qquad  & {\mbox{\sc Case $2$}}
   \label{rho2} \\
   {\hskip -1cm }  S(\rho_A) =   \frac {1} {6} \left [\;\ln{ \left (\frac {k^2} {16 k'}\right )} +
   \left( 2 - k^2 \right)  \frac {2 I(k) I(k')} {\pi} \right ] + \ln\;2,
   & &{\mbox{\sc Case $1$a}}
   \label{rho1a}
\end{eqnarray}
In our paper \cite{bik}, we have shown that equation
(\ref{3333May}) is valid for all three cases, which allowed us to
sum up the series  (\ref{3333May}) in case of  weak magnetic field
(including zero magnetic field) as well:
\begin{equation}
   {\hskip -2 cm} \clubsuit \qquad   S(\rho_A) =   \frac {1} {6} \left [\;\ln{ \left (\frac {k^2} {16 k'}\right )} + \left( 2 - k^2 \right)
         \frac {2 I(k) I(k')} {\pi} \right ] + \ln\;2,   \qquad \mbox{\sc Case $1$b},
   \label{rho1b}
\end{equation}
The  rigorous proof  and the precise history is given in the paper
\cite{big}. Here we report that these results are valid also for
$\gamma > 1$.

\begin{figure}
\begin{center}
\includegraphics[width=\columnwidth]{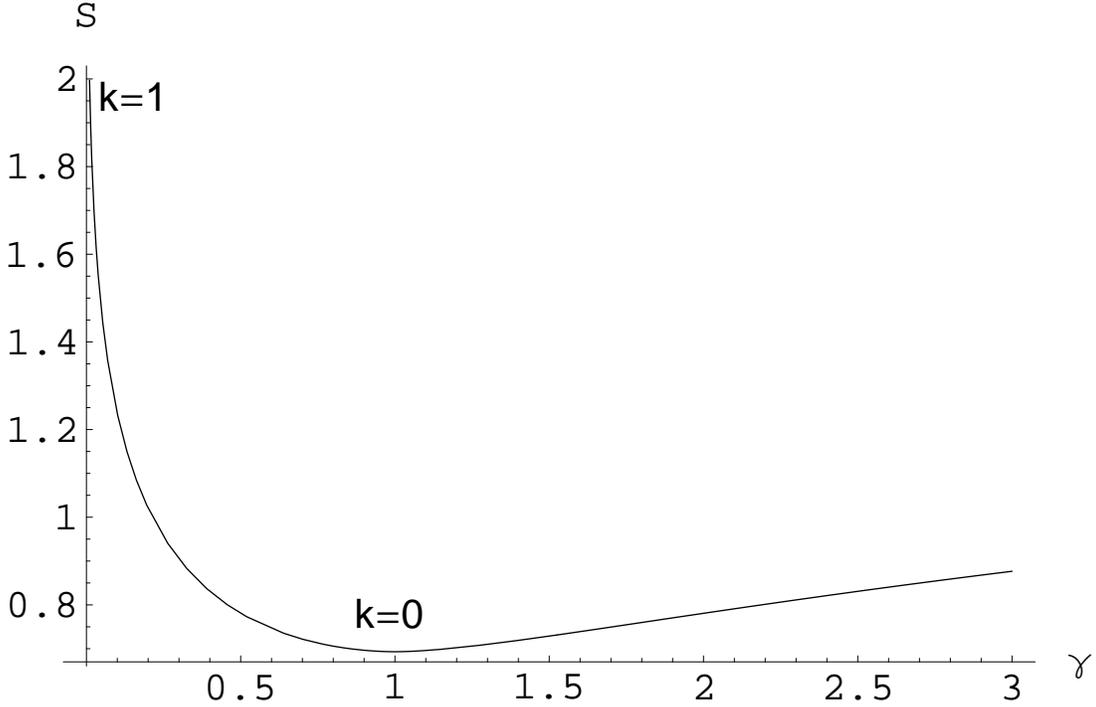}
\caption {The limiting entropy as a function of the anisotropy
parameter at constant vanishing magnetic field $h = 0$. The
entropy has a minimum $S=\ln2$ at $\gamma=1$ corresponding to the
boundary between cases {\sc $1$a} and {\sc $1$b}. $S$ diverges  to
$+ \infty$ at the phase transition $\gamma = 0$.}
   \label{h=0Plot}
\end{center}
\end{figure}

\begin{figure}
\begin{center}
\includegraphics[width=\columnwidth]{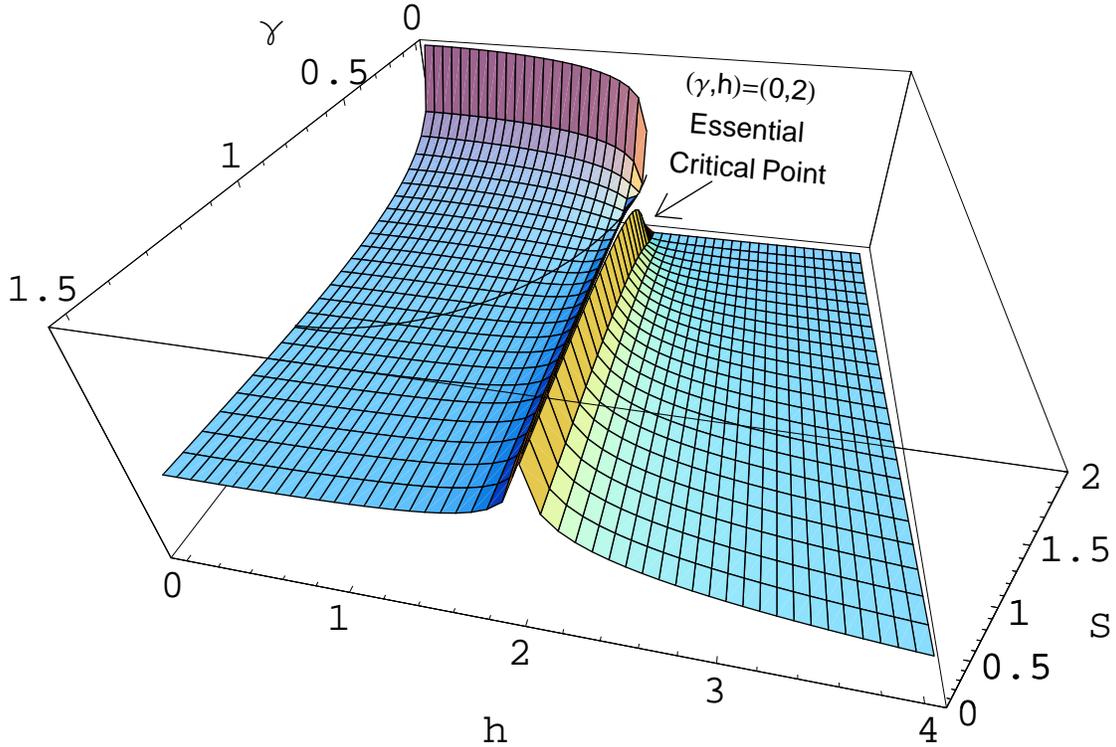}
\caption {Three-dimensional plot of the limiting entropy as a
function of the anisotropy parameter $\gamma$ and of the external
magnetic field $h$. The local minimum $S=\ln2$ at the boundary
between cases {\sc $1$a} and {\sc $1$b} is visible and marked by a
continuum line. $S$ diverges to $+ \infty$ at the phase
transitions $h = 2$ and $\gamma =0$, $h \le 2$. The entropy takes
every positive value in the vicinity of the essential critical
point $(h,\gamma)=(2,0)$.}
   \label{3DPlot}
\end{center}
\end{figure}

\begin{figure}
\begin{center}
\includegraphics[width=\columnwidth]{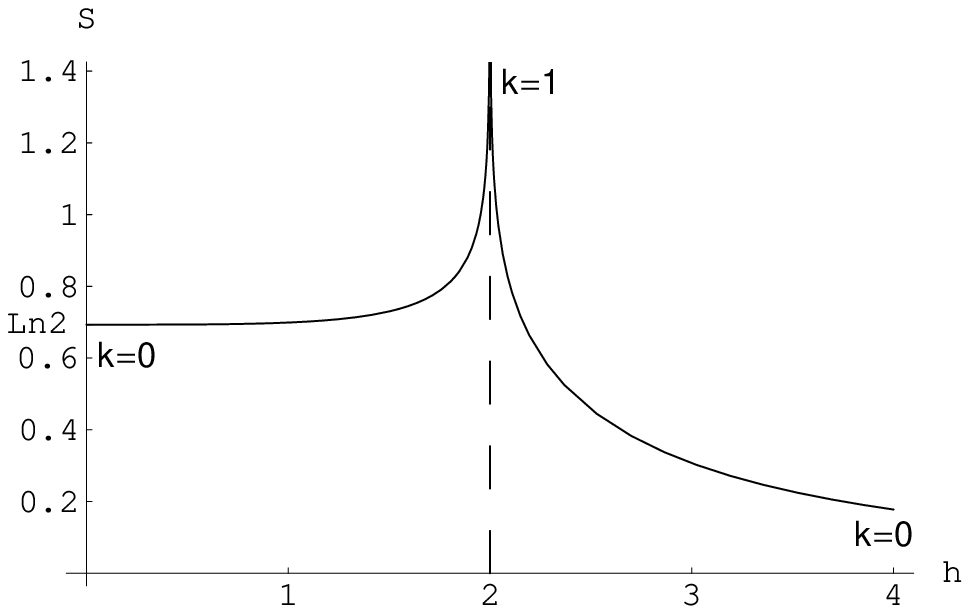}
\caption {The limiting entropy as a function of the magnetic field
at the Ising point $\gamma = 1$. The entropy has a local minimum
$S=\ln2$ at $h=0 $ and the absolute minimum for $h \to \infty$
where it vanishes. $S$ is singular at the phase transition $h=2$
where it diverges to $+ \infty$.}
   \label{IsingPlot}
\end{center}
\end{figure}

\section{Entropy's analysis}

Now we  can study the range of variation of the limiting entropy.
We find a {\bf  local minimum} $ S(\rho_A)=\ln2$ at the boundary
between cases {\sc $1$a} and {\sc $1$b} ($h=2\sqrt{1-\gamma^2}$).
This is the case of doubly degenerated ground state (\ref{deg})
and it is consistent with \cite{peschelemery}, where it was shown
that when the ground state becomes a superposition of two product
states with different quantum numbers, then the entropy of a
subsystem turns into $ \ln2$.

The absolute minimum ($S=0$) is achieved at infinite magnetic
field or at $\gamma =0$ for $h>2$, where the ground state becomes
ferromagnetic (i.e. all spins are parallel). The entropy diverges
to $+ \infty$, i.e has singularities, at the {\it phase
transitions}: $h=2$   or $\gamma=0$. To show this behavior of the
limiting entropy, we plot it as a function of the magnetic field
$h$ at constant anisotropy $\gamma = 1/2$ in
Fig.~\ref{entropyplot}.

We provide a plot of the entropy as a function of $\gamma$ at
constant (vanishing) magnetic field ($h=0$) in Fig.~\ref{h=0Plot},
where the local minimum $S = \ln 2$ is again visible at
$\gamma=1$. Figure \ref{3DPlot} is a three-dimensional plot of the
entropy as a function of both the anisotropy parameter $\gamma$
and of the external magnetic field $h$ and all the feature
discussed so far are visible.

{\bf The Ising point.} The degenerate product states case
(\ref{deg}) is particularly interesting. For $h=0$, $\gamma = 1$,
the $XY$ model reduces to the Ising model and the ground state
(\ref{deg}) is given by the {\it Bell-Pair} state:
\begin{eqnarray}
  |GS_1\rangle & = & \prod_{n\in \mbox{\rm lattice}} {1 \over \sqrt{2} } \Big( | \uparrow _n \rangle + (-1)^{n} | \downarrow _n \rangle \Big)
  \nonumber   \\
  |GS_2\rangle & = & \prod_{n\in \mbox{\rm lattice}} {1 \over \sqrt{2} } \Big( | \uparrow _n \rangle - (-1)^{n} | \downarrow _n \rangle \Big) .
  \label{BellPair}
\end{eqnarray}
We plot the entropy as a function of the magnetic field at the
Ising point $\gamma = 1$ in Fig.~\ref{IsingPlot}. One can notice
that the local minimum $S = \ln 2$ is achieved at $h = 0$.

\begin{figure}
   \dimen0=\textwidth
   \advance\dimen0 by -\columnsep
   \divide\dimen0 by 2
   \noindent\begin{minipage}[t]{\dimen0}
   \begin{flushleft}
   (a):
   \end{flushleft}
   \includegraphics[width=\columnwidth]{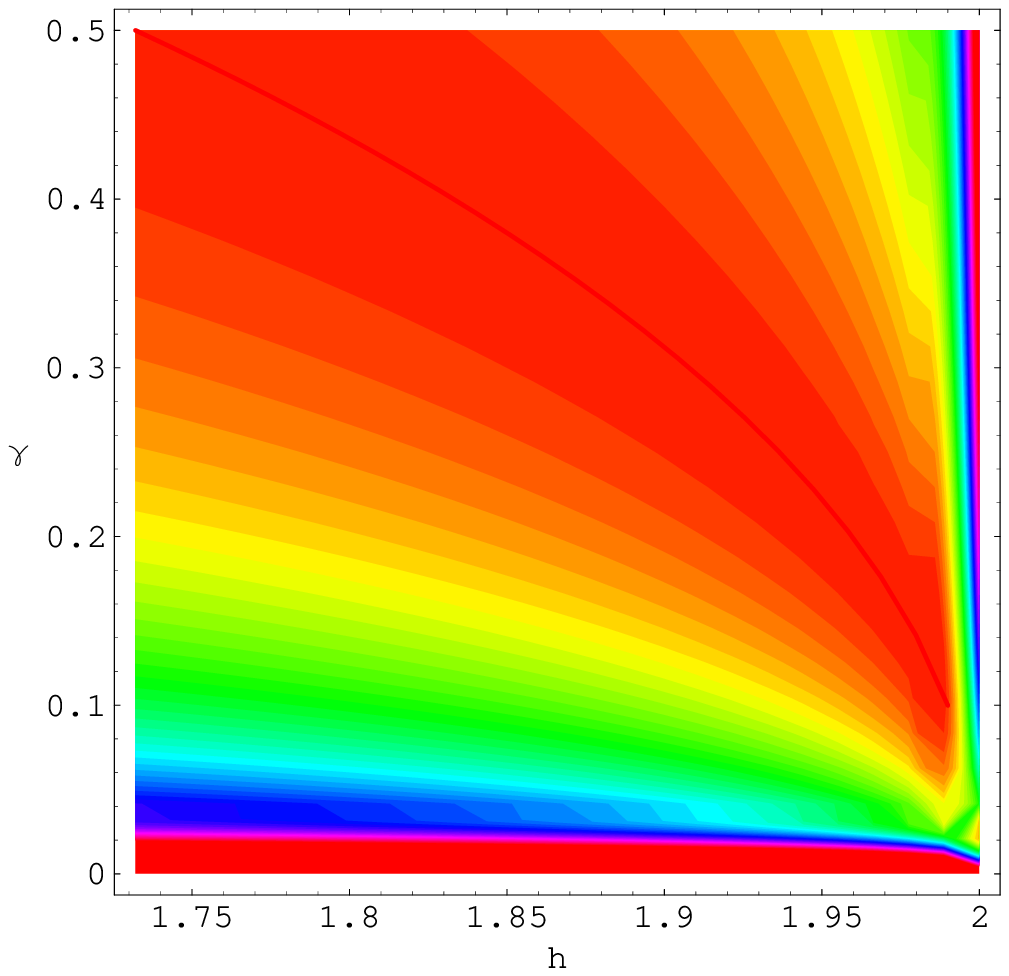}
   \end{minipage}
   \hfill
   \begin{minipage}[t]{\dimen0}
   \begin{flushleft}
   (b):
   \end{flushleft}
   \includegraphics[width=\columnwidth]{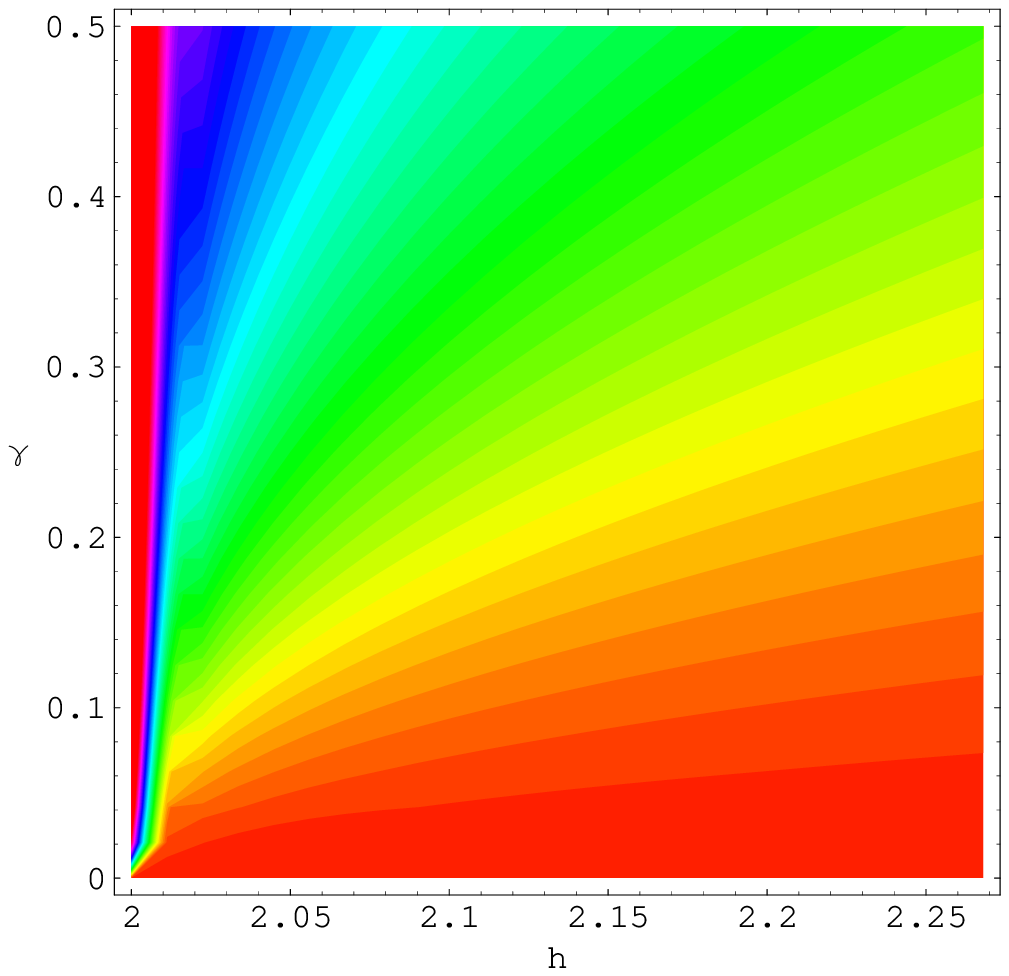}
   \end{minipage}
   \caption
   {Contour plot of the limiting entropy near the essential critical point $h=2$, $\gamma=0$.
   Regions of similar colors have similar entropy values and the lines where colors change are lines
   of constant entropy. $S (\rho_A)$ is diverging to $+ \infty$ on the critical lines $h=2$ and $h<2$, $\gamma=0$.
   One can see that near the essential critical point the lines of constant entropy grow denser.}
   \label{ContourPlot}
\end{figure}

{\bf The essential critical point.} Another interesting limit is
reached around the point $\gamma =0$, $h=2$. This point belongs to
both of the critical phases of the $XY$ model, so the entropy does
not  have an analytical expression [fixed value] on this point,
but we can study the behavior of $S (\rho_A)$ in vicinity of  this
point. We already studied a couple of trajectories reaching this
critical point: along the boundary  between cases {\sc $1$a} and
{\sc $1$b} ($h = 2 \sqrt{1 - \gamma^2}$) the entropy is on its
local minimum $S = \ln 2$. Along the critical lines ($h=2$ and
$\gamma = 0$ for $h<2$) the entropy is divergent, while for
$\gamma =0$ and $h>2$ the ground state is ferromagnetic and the
entropy is $0$. Since the limit of the entropy reaching the point
$(h,\gamma) =(2,0)$ does not exist (it is  direction-dependent),
we call this point {\it essential critical point}. In the next
section we study the vicinity of this point and show that
depending on the direction of approach the entropy can take any
positive value. In Fig.~\ref{ContourPlot} we present a contour
plot of the entropy around the essential critical point. From this
plot one can see that the entropy can assume a wide range of
values near the point.

\section{Ellipses and hyperbolas of constant entropy}

We now look for curves of constant entropy. Since the entropy
depends only on the elliptic parameter (\ref{modMay}), curves of
constant entropy are curves of constant $k$. Such trajectories are
easily found and the family of curves of constant entropy can be
written in terms of a single parameter $\kappa$:
\begin{eqnarray}
    \hskip -1.5cm
    \mbox{\sc Case $2$} \quad \: \: \bigg\{ \: h>2 \: : &
    \Bigg( { h \over 2 } \Bigg)^2 - \Bigg( {\gamma \over \kappa } \Bigg)^2 = 1,  \qquad &  0 \le \kappa <\infty
    \label{hyperbola} \\
    \hskip -1.5cm
    \mbox{\sc Case $1$a} \quad \Bigg\{ \small{
                               \begin{array}{l}
                               h<2, \\
                               \gamma > \sqrt{1 - (h/2)^2} \\ \end{array} } \Bigg. : \quad &
    \Bigg( { h \over 2 } \Bigg)^2 + \Bigg( {\gamma \over \kappa } \Bigg)^2 = 1,  \qquad & \kappa > 1
    \\
    \hskip -1.5cm
    \mbox{\sc Case $1$b} \quad \Bigg\{ \small{
                               \begin{array}{l}
                               h<2, \\
                               \gamma < \sqrt{1 - (h/2)^2} \\ \end{array} } \Bigg. : \quad &
    \Bigg( { h \over 2 } \Bigg)^2 + \Bigg( {\gamma \over \kappa } \Bigg)^2 = 1,  \qquad &  \kappa < 1 .
    \label{ellipse}
\end{eqnarray}
For $h>2$ (Case $2$), the curves of constant entropy are
hyperbolas, while for $h<2$ (Case {\sc $1$a} and {\sc $1$b}) they
are ellipses.

Each point in the phase diagram of the $XY$ model belongs to one
of these curves. By selecting a value of the parameter $\kappa$,
we select a family of point with the same elliptic parameter $k$
in (\ref{modMay}). There is a one-to-one correspondence between
$k$ and $\kappa$:
\begin{eqnarray}
   k = \sqrt{ {\kappa^2 \over 1 + \kappa^2} } \qquad \qquad &
   k' = \sqrt{ {1 \over 1 + \kappa^2 } } & \qquad \qquad \mbox{\sc Case $2$} \\
   k = \sqrt{ {\kappa^2 - 1 \over \kappa^2} } \qquad \qquad &
   k' = {1 \over \kappa } & \qquad \qquad \mbox{\sc Case $1$a} \\
   k = \sqrt{ 1 - \kappa^2 } \qquad \qquad & k' = \kappa & \qquad \qquad \mbox{\sc Case $1$b} .
\end{eqnarray}
We recognize that $\kappa=1$, as the boundary between Case {\sc
$1$a} and {\sc $1$b}, is the curve where the ground state can be
expressed as a doubly degenerate product state (\ref{deg}).

It is important to notice that this curves have all the essential
critical point $(h,\gamma)=(2,0)$ in common. This means that
starting from any point in the phase diagram of the $XY$ model,
one always reaches the essential critical point by following a
curve of constant entropy.

For $h<2$, the entropy has a minimum at $\ln 2$ and diverges to $+
\infty$ approaching the critical line $h=2$. For $h>2$, $S
(\rho_A)$ decreases monotonically from $+ \infty$ near the
critical line to $0$ at infinite magnetic field. Beside the
critical lines, the entropy is a continuous function, so its range
is the positive real axis.

This means that, depending on the direction of approach, the
entropy assumes every positive number near the essential critical
point, since every ellipse or hyperbola of constant entropy passes
through that point. In other words, a small deviation from the
essential critical point can bring a big change in the value of
the entropy. This is very important from the point of view of {\bf
quantum control}, because it allows to change dramatically the
entanglement (and hence the quantum computing capabilities) with
small changes in the parameters of the system.

It is easy to express the entropy in terms of the parameter
$\kappa$ defining the ellipses and hyperbolas of constant entropy:
\begin{eqnarray}
    \hskip -1.5cm
    S(\rho_A)=  \frac {1} {6} \left [\;\ln{ \frac {4 (\kappa^2 +1)} {\kappa} } +
    \frac {2} {\pi} \frac {\kappa^2-1} {\kappa^2 +1}
    I \left( \sqrt{ \frac {\kappa^2} {\kappa^2 +1} } \right) I \left( \sqrt{ \frac {1} {\kappa^2 + 1} } \right) \right ]
    && \quad \mbox{\sc Case $2$} \nonumber \\
    \hskip -1.5cm
    S(\rho_A) =  \frac {1} {6} \left [\;\ln{ \frac {\kappa^2 - 1} {16 \kappa} } +
    \frac {2} {\pi} \frac {(\kappa^2 + 1)} {\kappa^2}
    I \left( \sqrt{ \frac {\kappa^2 - 1} {\kappa^2} } \right) I \left( \frac {1} {\kappa} \right)
    \right ] + \ln\;2, &&  \quad \mbox{\sc Case $1$a} \nonumber \\
    \hskip -1.5cm
    S(\rho_A) =  \frac {1} {6} \left [\;\ln{ \frac {1 - \kappa^2} {16 \kappa} } +
    \frac {2} {\pi} (\kappa^2 + 1)
    I \left( \sqrt{ 1- \kappa^2 } \right) I \left( \kappa \right)
    \right ] + \ln\;2, &&  \quad \mbox{\sc Case $1$b} . \nonumber \\
    \label{kappaentropy}
\end{eqnarray}

\section{Entropy approaching the critical lines}

Using the formulae of the previous section, we are now in a
position to discuss the divergences of the entropy near the
critical phases. General results exist in these cases, based on a
conformal field theory approach \cite{cardy} and specific results
were derived for the isotropic case ($\gamma = 0$) in \cite{jin}.
We know that, in the double-scaling limit we are considering, the
entropy diverges logarithmically with the size of the block. The
coefficient of this logarithmical divergence can be calculated by
knowing the central charge of the corresponding conformal field
theory at the critical point \cite{cardy}.

Setting $\kappa =0$ or $\kappa = \infty$, the ellipses and
hyperbolas of constant entropy (\ref{ellipse}) collapse into the
critical lines, i.e. a vertical line at $\gamma = 0$ or a
horizontal line at $h =2$, respectively. Using
(\ref{kappaentropy}), we can study how the entropy diverges
approaching these lines.
Using {\sc Case $1$b} in
(\ref{kappaentropy}), we can take $\kappa \to 0$ to find
\begin{eqnarray}
   S (\kappa \to 0, h < 2) & \sim & - {1 \over 3} \; \ln \left( {\kappa  \over 2} \right) + \ldots
   \nonumber \\
   &  = & - {1 \over 3} \; \ln \left( {\gamma  \over 2} \right)
   + {1 \over 6} \; \ln \left[ 1 - (h/2)^2 \right] + \ldots
\end{eqnarray}
which is consistent with the results obtained in \cite{jin} for
the isotropic case ($\gamma = 0$).

We can investigate how the entropy approaches the critical line $h
= 2$ from below and from above. In the former case, we shall set
$\kappa \to \infty$ in {\sc Case $1$a} of (\ref{kappaentropy}):
\begin{eqnarray}
   S (\kappa \to \infty, h<2) & \sim &  {1 \over 3} \; \ln \left( {\kappa \over 2} \right) + \ldots
   \nonumber \\
   & = & - {1 \over 6} \; \ln \left[ 1 - (h/2)^2 \right] + {1 \over 3} \; \ln \left( { \gamma \over 2} \right) + \ldots
\end{eqnarray}
In the latter case, for a direction almost parallel to the
critical line $h = 2$, but slightly above it, we take $\kappa \to
\infty$ in {\sc Case $2$} of (\ref{kappaentropy}):
\begin{eqnarray}
   S (\kappa \to \infty, h>2) & \sim &  {1 \over 3} \; \ln \left( 4 \; \kappa \right) + \ldots
   \nonumber \\
   & = & - {1 \over 6} \; \ln \left[ (h/2)^2 - 1 \right] + {1 \over 3} \; \ln \left( 4 \; \gamma \right) + \ldots
\end{eqnarray}
These results are in agreement with the conclusion of
\cite{cardy}.

\begin{table}
   \hskip -.7cm
   \begin{tabular}{|l|l|c|c|}
     \hline
      Region & $\qquad \qquad \qquad S (\rho_A)$  & \TT $\begin{array}{c}
      {\rm Curves \: of} \\ {\rm Constant} \: S \\
      \end{array} $ & $\begin{array}{c} {\rm Range \: of} \\ {\rm Parameters} \\
      \end{array} $ \BB \\
     \hline
     \hline
      {\sc $2$:}  $\quad h > 2$  & {\large
      $\frac {1} {6} \left [\;\ln{ \frac {4} {k \; k'}} +  \frac {2 (k^2-k'^2) I(k) I(k')} {\pi} \right ]$ } &
      $\left( { h \over 2 } \right)^2 - \left( {\gamma \over \kappa } \right)^2 = 1 $  & \T
      $\begin{array}{c}
      0 \le k < 1 \\ 0 \le \kappa < \infty \\ k = \sqrt{ {\kappa^2 \over 1 + \kappa^2} } \\
      \end{array} $ \B \\
     \hline
      {\sc $1$a}:  $\left\{ \small{ \begin{array}{l} h<2, \\ \gamma > \sqrt{1 - (h/2)^2} \\ \end{array} } \right. $  &
      $ \frac {1} {6} \left [\;\ln{ \frac {k^2} {16 k'} } +  \frac {2 ( 2 - k^2) I(k) I(k')} {\pi} \right ] + \ln\;2$ &
      $\left( { h \over 2 } \right)^2 + \left( {\gamma \over \kappa } \right)^2 = 1$ & \T
      $\begin{array}{c}
      0 < k < 1 \\ \kappa > 1 \\ k = \sqrt{ {\kappa^2 - 1 \over \kappa^2} } \\
      \end{array} $ \B \\
     \hline
      {\sc $1$b}:  $\left\{ \small{ \begin{array}{l} h<2, \\ \gamma < \sqrt{1 - (h/2)^2} \\ \end{array} } \right.$  &
      $ \frac {1} {6} \left [\;\ln{ \frac {k^2} {16 k'} } +  \frac {2 ( 2 - k^2 ) I(k) I(k')} {\pi} \right ] + \ln\;2 $ &
      $\left( { h \over 2 } \right)^2 + \left( {\gamma \over \kappa } \right)^2 = 1 $ & \TBT
      $\begin{array}{c}
      0 < k < 1 \\ \kappa < 1 \\ k = \sqrt{ 1 - \kappa^2 } \\
      \end{array} $ \BTB \\
     \hline
      $ \qquad \gamma = \sqrt{1 - (h/2)^2}$ & {\large$ \quad  \ln\;2 $  } &
      $\left( { h \over 2 } \right)^2 + \gamma^2 = 1 $ & \TT
      $\begin{array}{c}
      k = 0 \\ \kappa = 1 \\
      \end{array} $ \BB  \\
     \hline
   \end{tabular}
   \caption{Recap of the results with the entropy in the different regions of the phase diagram,
   the curves (ellipses and hyperbolas) of constant entropy and the relationship between
   the elliptic parameter $k$ and the parameter $\kappa$ defining the family of curves.}
\end{table}

\section{Conclusions}

We analyzed the entanglement in the ground state of the $XY$ model
on the infinite chain by studying the von Neumann entropy $S
(\rho_A)$ of a block $A$ of neighboring spins. This entropy is an
effective measure of the quantum computing capabilities of a
system and plays a fundamental role in the field of quantum
information.

Using previously known results for the entropy in the limit of a
large block of spins \cite{big}-\cite{bik}, we studied the
behavior of $S (\rho_A)$ in the phase diagram of the $XY$ model
(see Table~1). We found that for $h<2$, the entropy has a local
minimum $S = \ln 2$ on the curve $(h/2)^2 + \gamma^2 = 1$. On this
line the ground state is a doubly degenerate linear combination of
product states. The entropy diverges to $+ \infty$ at the phase
transitions $h=2$ and $\gamma =0$, $h<2$. For $h>2$, the entropy
reaches the absolute minimum at infinite magnetic field $h \to +
\infty$ and for $\gamma =0$, i.e. when the ground state is
ferromagnetic. $S (\rho_A)$ diverges to $+ \infty$ on the critical
line $h=2$ and it is continuous otherwise.

We identified a set of curves (ellipses and hyperbolas) of
constant entropy. They are given in
(\ref{hyperbola}-\ref{ellipse}). All these curves have one point
in common, that we decided to call {\it essential critical point}:
$(h,\gamma) = (2,0)$. The fact that all the curves of constant
entropy pass through one point, together with the fact that the
range of the entropy as a function of $\gamma$ and $h$ is the
positive real axis, means  that the entropy can assume any real
positive value near the essential critical point, depending on the
direction of approach. In turn, this means that the essential
critical point is very important for quantum control, in that
small changes in the parameters can change the entanglement
dramatically.

With this work, we conclude the analysis of the asymptotic Von
Neumann entropy for the bi-partite one-dimensional $XY$ model. We
covered the whole phase diagram (including $\gamma
> 1$), focusing on the sector $h \ge 0$ and $\gamma \ge 0$: since
the model is invariant under the substitution $\gamma \to -
\gamma$ or $h \to - h$ the results for the entanglement can be
extended immediately to negative values of the anisotropy
parameter $\gamma$ or of the magnetic field $h$.

Finally we note that the work done so far on the $XY$ model in a
constant magnetic field allows us to calculate the bipartite
entropy of the $XY$ model in a staggered magnetic field as well.
As we discuss in the appendix, there is exact mapping between
these two models. Therefore, the knowledge of the entanglement for
one of the models automatically gives the entanglement for the
other. We give some details in the appendix.

\section*{Acknowledgments}
 We would like to thank A. Abanov, P.Deift, B.McCoy,
I.Peschel and H.Widom for useful discussions. This work was
supported by NSF Grants DMS 0503712, DMR-0302758 and DMS-0401009.

\hfil

\appendix
\section{$XY$ model in a staggered magnetic field}

It is a well known fact in the theory of integrable models, that
there is an exact mapping between the traditional $XY$ model in
constant magnetic field described by the Hamiltonian (\ref{xxh})
and the $XY$ model in staggered magnetic field:
\begin{equation}
  \hskip -1cm
  {\cal H'} = - J \sum_{n=-\infty}^{\infty}
  (1+\gamma')\sigma^x_{n}\sigma^x_{n+1}+(1-\gamma')\sigma^y_{n}\sigma^y_{n+1} +
  (-1)^n \; h' \: \sigma^z_{n}.
  \label{xxsh}
\end{equation}

This mapping is achieved by performing a rotation of every other
spin along the $x$ direction. To identify the two Hamiltonians,
one also needs to substitute $\gamma \to 1/ \gamma$\footnote{Note
that the large anisotropy regime is so mapped into the small
anisotropy regime and vice-versa.} and to rescale the magnetic
field and the Hamiltonian by a factor of $1 / \gamma$ and $\gamma$
respectively:
\begin{equation}
    \gamma' = 1 / \gamma \qquad \qquad \qquad
    h' = h / \gamma \qquad \qquad \qquad
    J = \gamma .
    \label{mapp}
\end{equation}

In the main body of this article we analyzed the entanglement of
the $XY$ model in a constant magnetic field (\ref{xxh}). The
results we derived can be applied directly to calculate the
bi-partite entanglement of the $XY$ model in a staggered field
(\ref{xxsh}). All formulae are valid and to calculate the entropy
for a staggered field one only needs to take the appropriate
result and perform the substitutions (\ref{mapp}).

Using (\ref{mapp}), the spectrum of the $XY$ model in a staggered
magnetic field is:
\begin{equation}
   \epsilon_k = \sqrt{ \left( \gamma' \cos k - h' / 2 \right)^2 + \sin^2 k } .
   \label{sspectrum}
\end{equation}
From this, we see that the critical phase $h=2$ is mapped to the
line $h'=2 \gamma'$.

\begin{figure}
\begin{center}
   \dimen0=\textwidth
   \advance\dimen0 by -\columnsep
   \divide\dimen0 by 2
\includegraphics[width=\columnwidth]{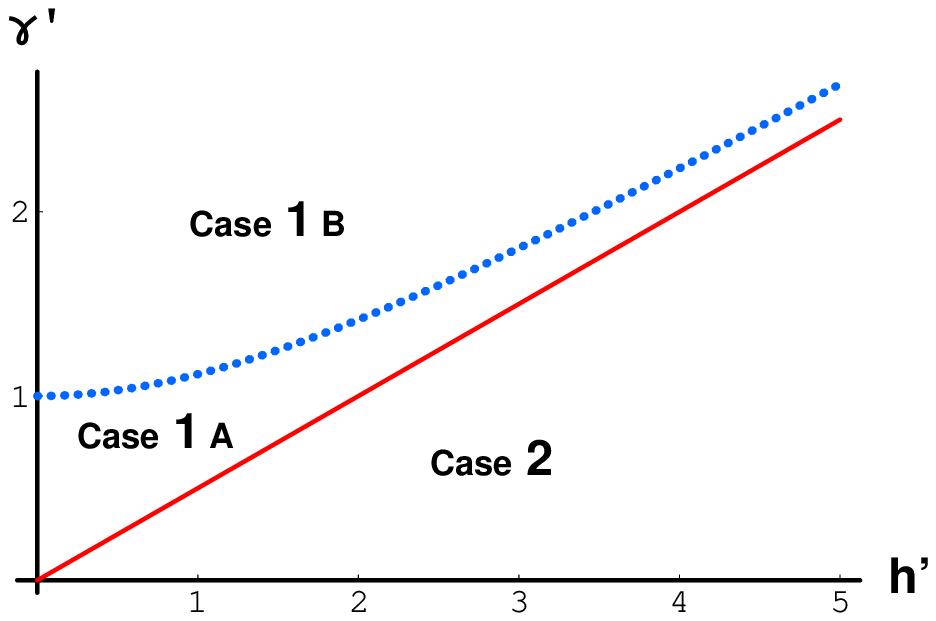}
\caption {Phase diagram of the anisotropic $XY$ model in a
staggered magnetic field (only $\gamma' \ge 0$ and $h' \ge 0$
shown). The three cases {\sc $2$, $1$a, $1$b}, considered in this
paper, are clearly marked. The critical phase $h' = 2 \gamma'$) is
drawn as bold line (red, online). The boundary between cases {\sc
$1$a} and {\sc $1$b}, where the ground state is given by two
degenerate product states, is shown as a dotted line (blue,
online).}
   \label{staggeredpd}
\end{center}
\end{figure}

Therefore, the mapping of the different cases and the definitions
of the elliptic parameter for this model are:
\begin{eqnarray}
   \hskip -1cm
   \mbox{\sc Case $2$} \qquad & \bigg\{ \: h' > 2 \gamma' \: :& \qquad
   k \equiv { 1 / \sqrt{ \left( h' / 2 \right) - \gamma'^2 + 1} }
   \label{stag2} \\
   \hskip -1cm
   \mbox{\sc Case $1$a} \qquad & \Bigg\{ \small{
                               \begin{array}{l}
                               h' < 2 \gamma', \\
                               \gamma' < \sqrt{1 + (h'/2)^2} \\ \end{array} } \Bigg. :& \qquad
   k \equiv \sqrt{ \left( h' / 2 \right) - \gamma'^2 + 1}
   \label{stag1a} \\
   \hskip -1cm
   \mbox{\sc Case $1$b} \qquad & \Bigg\{ \small{
                               \begin{array}{l}
                               h' < 2 \gamma', \\
                               \gamma' > \sqrt{1 + (h'/2)^2} \\ \end{array} } \Bigg. : & \qquad
   k \equiv { \sqrt{ \gamma'^2 - (h'/2)^2 -1 } \over \sqrt{ \gamma'^2 - (h'/2)^2} } \: .
   \label{stag1b}
\end{eqnarray}

We draw the phase diagram of the $XY$ model in a staggered field
and indicate the three cases in Fig.~\ref{staggeredpd}. With the
definitions (\ref{stag2}-\ref{stag1b}), one can plug the elliptic
parameter into (\ref{rho2}-\ref{rho1b}) and use the other results
of this paper to calculate the entropy of the $XY$ model in a
staggered magnetic field.

\hfil

\end{document}